\documentclass[journal]{IEEEtran}
\ifCLASSINFOpdf
\else
\fi
\usepackage{amsmath,amssymb}

\usepackage{graphicx}

\begin{document}
%
\title{Performance of SOI Pixel Sensors Developed for X-ray Astronomy}
%
%
%

\author{Takaaki~Tanaka, Takeshi~Go~Tsuru, Hiroyuki~Uchida, Sodai~Harada, Tomoyuki~Okuno, Kazuho~Kayama, Yuki~Amano, Hideaki~Matsumura, Ayaki~Takeda, Koji~Mori, Yusuke~Nishioka, Kohei~Fukuda, Takahiro~Hida, Masataka~Yukumoto, Yasuo~Arai, Ikuo~Kurachi, Shoji~Kawahito, Keiichiro~Kagawa, Keita~Yasutomi, Sumeet~Shrestha, Syunta~Nakanishi, Hiroki~Kamehama, Takayoshi~Kohmura, Kouichi~Hagino, Kousuke~Negishi, Kenji~Oono, and Keigo~Yarita
\thanks{T.~Tanaka, T.~G.~Tsuru, H.~Uchida, S.~Harada, T.~Okuno, K.~Kayama, and Y.~Amano are with Kyoto University.}
\thanks{H.~Matsumura is with University of Tokyo}
\thanks{A.~Takeda, K.~Mori, Y.~Nishioka, K.~Fukuda, T.~Hida, and M.~Yukumoto are with University of Miyazaki}
\thanks{Y.~Arai and I.~Kurachi are with KEK}
\thanks{S.~Kawahito, K.~Kagawa, K.~Yasutomi, S.~Shrestha, and S.~Nakanishi are with Shizuoka University}
\thanks{H. Kamehama is with National Institute of Technology, Okinawa College}
\thanks{T.~Kohmura, K.~Hagino, K.~Negishi, K.~Oono, and K.~Yarita are with Tokyo University of Science}
\thanks
}

\maketitle

\begin{abstract}
We have been developing monolithic active pixel sensors for X-rays based on the silicon-on-insulator technology. Our device consists of a low-resistivity Si layer for readout CMOS electronics, a high-resistivity Si sensor layer, and a SiO$_2$ layer between them. This configuration allows us both high-speed readout circuits and a thick (on the order of {\boldmath $100~\mu{\rm m}$}) depletion layer in a monolithic device. Each pixel circuit contains a trigger output function, with which we can achieve a time resolution of {\boldmath $\lesssim 10~\mu{\rm s}$}. One of our key development items is improvement of the energy resolution. We recently fabricated a device named XRPIX6E, to which we introduced a pinned depleted diode (PDD) structure. The structure reduces the capacitance coupling between the sensing area in the sensor layer and the pixel circuit, which degrades the spectral performance.  With XRPIX6E, we achieve an energy resolution of \boldmath$\sim 150$~eV in full width at half maximum for 6.4-keV X-rays. In addition to the good energy resolution, a large imaging area is required for practical use. We developed and tested XRPIX5b, which has an imaging area size of {\boldmath $21.9~{\rm mm} \times 13.8~{\rm mm}$} and is the largest device that we ever fabricated. We successfully obtain X-ray data from almost all the {\boldmath $608 \times 384$} pixels with high uniformity. 
\end{abstract}


%

\section{Introduction}
\begin{figure}[hb]
\centering
\includegraphics[width=1.0\linewidth]{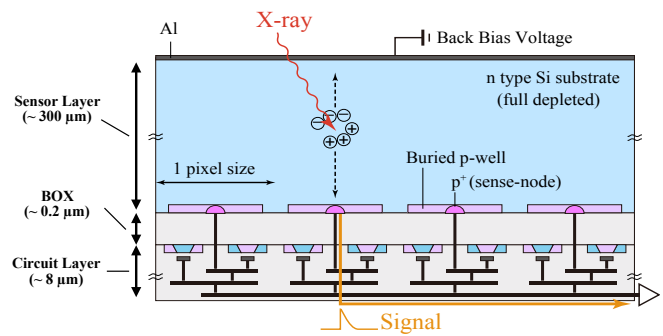}
\caption{Cross-sectional view of our SOI pixel sensor.}
\label{fig:cross}
\end{figure}

%
%
%
%
\IEEEPARstart{W}{e} have been developing active pixel sensors based on the silicon-on-insulator (SOI) technology, whose 
schematic picture is shown in Fig.~\ref{fig:cross}. 
Our device consists of a low-resistivity Si layer for readout CMOS electronics, a high-resistivity Si sensor layer, and a SiO$_2$ layer in between. 
This configuration allows us both high-speed readout circuits and a thick depletion layer in a monolithic device. 
In a series of devices we have been developing for X-ray detection, named XRPIX \cite{tsuru2014}, we implement trigger output circuits in each pixel, with 
which we can achieve a time resolution of $\lesssim 10~\mu{\rm s}$. 
This provides a huge advantage over X-ray charge-coupled devices \cite{tanaka2018}--\cite{turner2001}, which are the de facto standard as X-ray sensors and typically have 
a time resolution of $\sim {\rm s}$. 
The superior time resolution enables us not only to detect short-timescale phenomena but also to employ anti-coincidence shields. 
The anti-coincidence shield provides lower instrumental backgrounds, which are essential to achieve high sensitivity in various applications including X-ray astronomy. 

One of the applications of XRPIX we are aiming at is focal plane detectors for the FORCE (Focusing On Relativistic universe and Cosmic Evolution) mission \cite{mori2016}, 
a future Japan-led X-ray astronomy satellite. 
FORCE covers a wide bandpass of 1--80~keV with a fine angular resolution of $< 15~{\rm arcmin}$ in a half-power diameter. 
In the baseline design, the focal plane detector consists of two sensors stacked together: an SOI pixel sensor in the top layer for lower energy X-rays at  $\lesssim 20~{\rm keV}$ and a CdTe double-sided strip detector \cite{watanabe2009} at the bottom for higher energy X-rays. 
The requirements for the SOI pixel sensor are a pixel size smaller than $200~\mu{\rm m} \times 200~\mu{\rm m}$, an energy resolution better than 300~eV in 
full width at half maximum (FWHM) at 6~keV, and an imaging area size larger than $20~{\rm mm} \times 20~{\rm mm}$. 
Our devices typically have a pixel size of $36~\mu{\rm m} \times 36~\mu{\rm m}$ and fulfill the first requirement. 
Our recent development focuses on improvement of the energy resolution and on fabrication of large-sized sensors. 

Here we report on performance of our recently fabricated devices: XRPIX6E and XRPIX5b. 
We introduce a new pixel structure to XRPIX6E for better spectral performance, and thus we present 
our test results focusing on its spectral performance. The other device, XRPIX5b, is the largest sensor 
that we have ever fabricated. We mainly evaluate its gain uniformity.

\section{Results from XRPIX6E}
\subsection{Interference and Charge Collection Efficiency in XRPIX}
Spectral performance has been one of our key development items. 
The best spectral performance so far is obtained with XRPIX3b \cite{takeda2015}, in which 
we implement charge sensitive amplifiers (CSAs) as the first-stage amplifiers of the pixel circuit 
instead of source followers used in our earlier devices. 
Replacing the source followers with the CSAs, we succeeded in increasing the gain by a factor of 3.3 from 
5.4~$\mu{\rm V}/e^-$ to 17.8~$\mu{\rm V}/e^-$ \cite{takeda2015}. 
Thanks to the increased gain, we achieved an energy resolution of 320~eV (FWHM) for 6-keV X-rays \cite{takeda2015}. 
However, the spectral performance of XRPIX3b reaches this level only when we read out it in the frame readout 
mode, in which we serially read out all the pixels without using the trigger function. 
The spectral performance is considerably degraded when we read out it in the event-driven readout mode, in which 
we read out a triggering pixel and its surrounding pixels upon a trigger output \cite{tsuru2018}. 
Also, output signals have a large offset in the event-driven readout mode \cite{takeda2014}. 
We found that these are caused by a capacitive coupling between the buried p-well (BPW) in the sensor layer (Fig.~\ref{fig:cross}) and the 
trigger signal line in the circuit layer \cite{tsuru2018,takeda2014}. 
Therefore, reduction of the capacitive coupling is a key to achieving better spectral performance particularly in the event-driven readout mode. 
We employed the Double-SOI structure in a sensor named XRPIX3-DSOI, and successfully reduced the capacitive coupling \cite{ohmura2016}. 
As detailed below, we here tried another new structure for the same purpose. 

Charge collection efficiency (CCE) should be high enough to achieve superior spectral performance. 
We irradiated XRPIX1b and XRPIX3b with an X-ray pencil beam, and 
found that the CCE becomes lower when X-rays are absorbed near the pixel boundaries \cite{matsumura2015,negishi2018}. 
Comparing the experimental data with results from device simulations that we carried out, we came to a conclusion that 
a part of signal charge is lost due to traps located close to the interface between the sensor and SiO$_2$ layers \cite{matsumura2015}. 
The results indicate that the pixel structure need to be redesigned in order to avoid signal charge from being trapped at the interface.

\subsection{Specifications of XRPIX6E}
\begin{figure}[ht]
\centering
\includegraphics[width=1.0\linewidth]{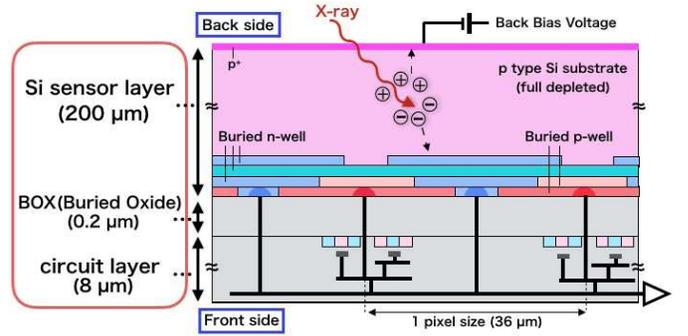}
\caption{Cross-sectional view of XRPIX6E.}
\label{fig:6e_cross}
\end{figure}

\begin{figure}[ht]
\centering
\includegraphics[width=0.7\linewidth]{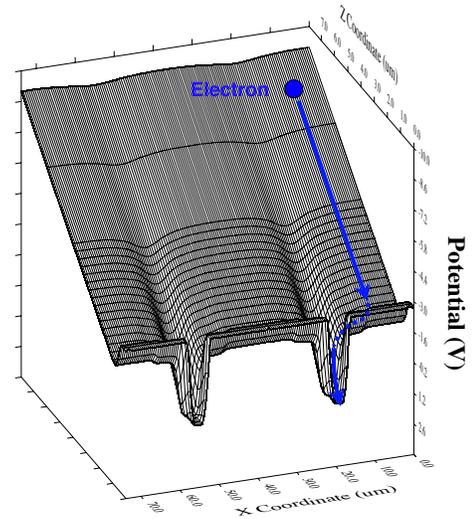}
\caption{Simulated electric potential in the sensor layer of XRPIX6E.}
\label{fig:6e_potential}
\end{figure}

In order to overcome the aforementioned issues, we developed a new device, XRPIX6E \cite{harada2018}. 
A schematic cross sectional view of XRPIX6E is presented in Fig.~\ref{fig:6e_cross}. 
We employed the pixel structure called pinned depleted diode, which was recently developed by Kamehama {\it et al.} \cite{kame2018}. 
The highly doped BPW, to which we apply a fixed voltage, acts as an electrostatic shield, and thereby is expected to reduce the 
capacitive coupling between the sensor and circuit layers. 
The step-like buried n-wells (BNWs) help signal charge be drifted toward the sense node without touching the interface between the 
sensor and SiO$_2$ layers as illustrated in Fig.~\ref{fig:6e_potential}. 
We thus expect that the CCE of XRPIX6E is higher than that of our previous devices with the conventional pixel structure shown in Fig.~\ref{fig:cross}. 

Except for the pixel structure, the specifications of XRPIX6E  are basically similar to our recent devices in the XRPIX series. 
The sensor layer was fabricated on a p-type float zone wafer with a typical resistivity of $> 25~\Omega~{\rm cm}$, and has a thickness of $200~\mu{\rm m}$. 
XRPIX6E has $48 \times 48$ pixels with a pixel size of $36~\mu{\rm m} \times 36~\mu{\rm m}$. 
The pixel circuit has a CSA followed by a correlated double sampling circuit. 
The signal is further processed by the peripheral readout circuit composed of a column amplifier and an output buffer. 

\begin{figure*}[tb]
\centering
\includegraphics[width=1\linewidth]{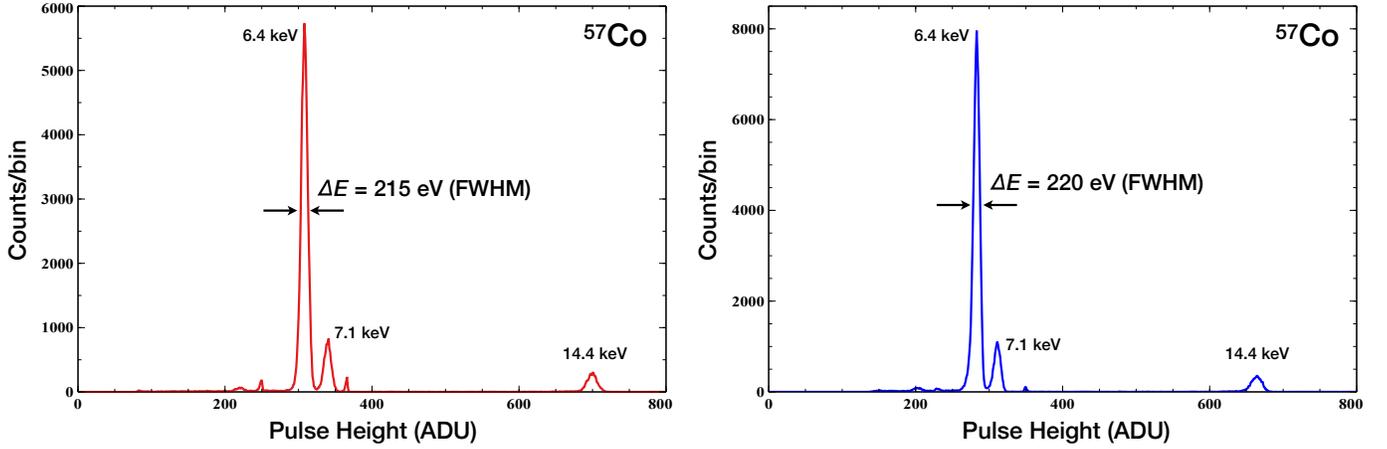}
\caption{Spectra of X-rays/$\gamma$-rays from $^{57}{\rm Co}$ obtained with central $8 \times 8$ pixels of XRPIX6E. The left and right panels show spectra obtained in the frame readout mode and the event-driven readout mode, respectively.}
\label{fig:6e_spectra}
\end{figure*}

\subsection{Spectral Performance of XRPIX6E}
In order to examine the spectral performance, 
we irradiated XRPIX6E with X-rays/$\gamma$-rays from $^{57}{\rm Co}$ (Fe K$\alpha$ at 6.4~keV, Fe K$\beta$ at 7.1~keV, and $\gamma$-rays at 14.4~keV), 
and read out central $8 \times 8$ pixels of XRPIX6E in the frame readout mode and in the event-driven readout mode. 
We applied a back-bias voltage of $-200~{\rm V}$ to the sensor layer so that it is fully depleted. 
The measurements were carried out at a temperature of $-60~^\circ{\rm C}$. 
In the frame readout mode, we search for a pixel whose output pulse hight exceeds a preset threshold (event threshold) in each frame. 
We then examine outputs from surrounding 8 pixels to see whether or not any of them exceeds a threshold for charge sharing (split threshold). 
Likewise, in the event-driven readout mode, we examine 8 pixels surrounding the triggering pixels. 
Events are classified into single-pixel events, double-pixel events, triple-pixel events, etc. depending on the hit patterns. 
In what follows, we show spectra of single-pixel events, in which outputs from all the surrounding 8 pixels are lower than the split threshold. 

\begin{figure}[ht]
\centering
\includegraphics[width=1\linewidth]{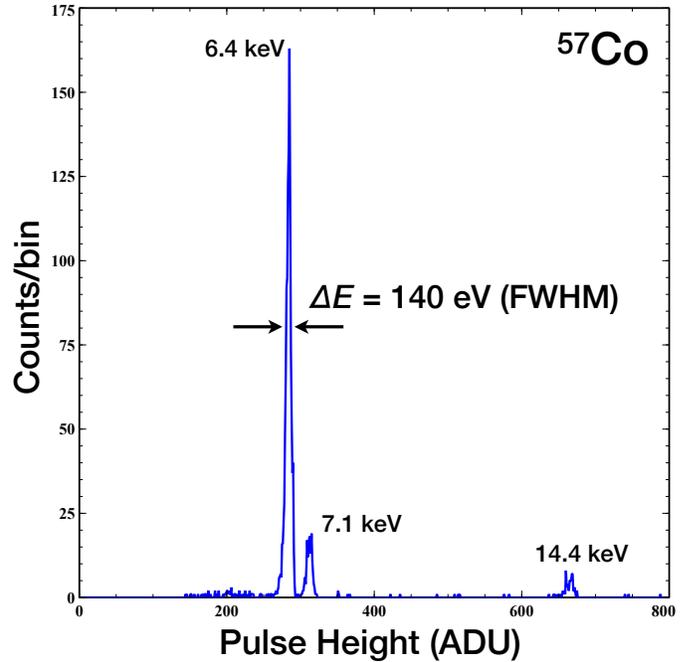}
\caption{Spectrum of X-rays/$\gamma$-rays from $^{57}{\rm Co}$ from the best pixel of XRPIX6E operated in the event-driven readout mode.}
\label{fig:6e_best_spectum}
\end{figure}

We show spectra obtained with XRPIX6E in Fig.~\ref{fig:6e_spectra}. Note that correction of the pixel-to-pixel gain variation is not applied. 
The energy resolutions at 6.4~keV are 215~eV and 220~eV (FWHM) in the frame readout mode and the event-driven readout mode, respectively, 
which are the best values ever achieved with our XRPIX devices. 
If we pick up the best pixel in the $8 \times 8$ pixels that we read out, we obtained the spectrum presented in Fig.~\ref{fig:6e_best_spectum}. 
The energy resolution is 140~eV (FWHM) for 6.4-keV X-rays. 
An important fact to note is that the spectral resolutions in the two modes are comparable to each other, which indicates that 
the capacitive coupling between the sensor and circuit layers is substantially reduced, compared to XRPIX3b. 
Output signals in the event-driven readout mode do not have a large offset any more \cite{harada2018}, indicating again reduced interference. 
We thus conclude that the highly doped BPW (Fig.~\ref{fig:6e_cross}) is effectively working as an electrostatic shield. 
The reduced interference would have resulted in readout noise reduction and thus in the good energy resolution in the event-driven readout mode. 
Another thing to note is that the energy resolution is improved even in the frame readout mode. 
The improvement would be ascribed to the increased gain of XRPIX6E, which was measured to be 46--48~$\mu{\rm V}/e^-$ \cite{harada2018}. 
The gain is by a factor of $\sim 2$ larger than that of XRPIX3b although the two devices have CSAs with the same design. 
The capacitive coupling between the sensor and circuit layers made the feedback capacitance of the CSA effectively smaller in XRPIX3b. 
In the XRPIX6E, on the other hand, the effective feedback capacitance has become larger thanks to the PDD structure working as a shield.

We are now evaluating the sub-pixel response of XRPIX6E in order to verify the CCE indeed is improved with the PDD structure. 
The result will be reported elsewhere. 

\section{Results from Large-Sized Device, XRPIX5b}
\subsection{Specifications of XRPIX5b}
\begin{figure}[ht]
\centering
\includegraphics[width=0.8\linewidth]{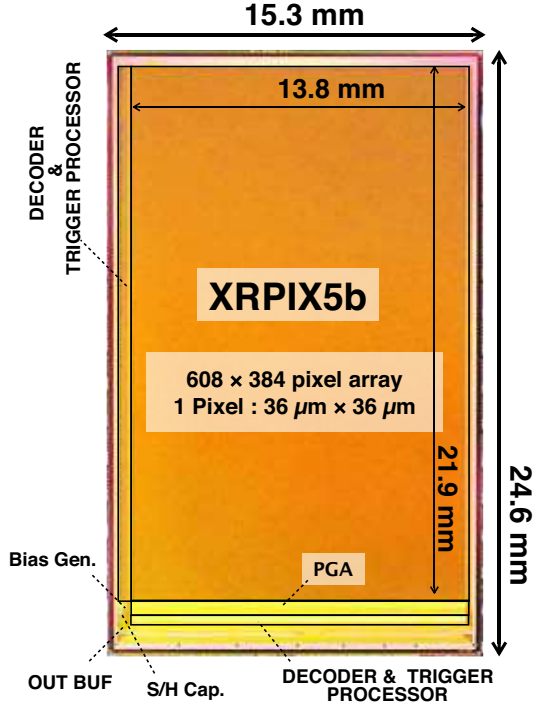}
\caption{Photograph of XRPIX5b.}
\label{fig:5b_photo}
\end{figure}

XRPIX5b \cite{hayashi2018} is the largest device in the XRPIX series that we have ever processed. 
A photograph of XRPIX5b is shown in Fig.~\ref{fig:5b_photo}. 
The imaging area has a size of $21.9~{\rm mm} \times 13.8~{\rm mm}$, where pixels 
with a size of $36~\mu{\rm m} \times 36~\mu{\rm m}$ are arranged in a $608 \times 384$ array. 
Since XRPIX5b is a three-sided buttable sensor, we can fulfill the requirement on the imaging area size 
for the FORCE mission ($20~{\rm mm} \times 20~{\rm mm}$) by tiling two chips in a $1 \times 2$ array. 
The pixel structure is the conventional one shown in Fig.~\ref{fig:cross}. 
The sensor layer, fabricated on a n-type float zone wafer, has a thickness of $300~\mu{\rm m}$. 
The pixel circuit is almost the same as that of XRPIX3b \cite{takeda2015}.

\begin{figure}[ht]
\centering
\includegraphics[width=1.0\linewidth]{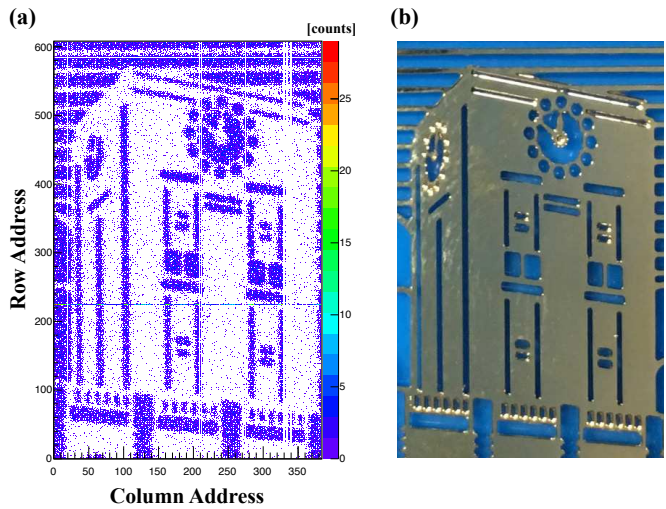}
\caption{(a) X-ray shadow image obtained with XRPIX5b illuminated with Ag K$\alpha$ (22~keV) and K$\beta$ (25~keV) lines from $^{109}{\rm Cd}$. (b) Photograph of the metal mask used to take the X-ray shadow image.} 
\label{fig:5b_image}
\end{figure}

In order to demonstrate its imaging capability, XRPIX5b, with a metal mask placed right above it, was illuminated by 
X-rays from a $^{109}{\rm Cd}$ source (Ag K$\alpha$ at 22 keV and Ag K$\beta$ at 25 keV). 
In Fig.~\ref{fig:5b_image}, we show the X-ray shadow image as well as a photograph of the metal mask. 
We operated XRPIX5b in the event-driven readout mode at a room temperature. 
The back-bias voltage applied was 10~V. 
The event rate during the measurement was $\sim 60$~Hz. 
We note that, in another measurement using an $^{241}{\rm Am}$ source, we achieved a throughput of $\sim 570$~Hz with XRPIX5b.

\subsection{Gain Uniformity}
\begin{figure}[th]
\centering
\includegraphics[width=0.9\linewidth]{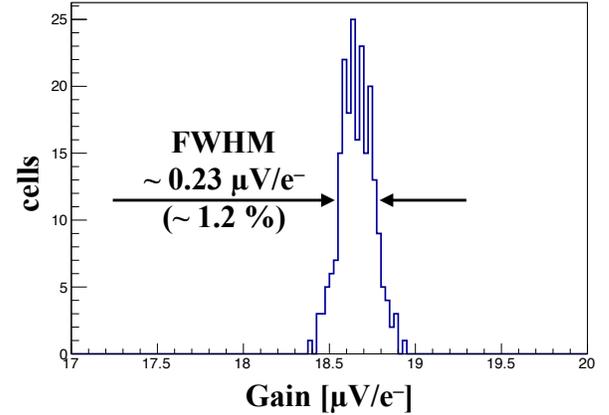}
\caption{Distribution of gains of $32 \times 32$ pixel islands (cells) of XRPIX5b.} 
\label{fig:5b_gain_dist}
\end{figure}

\begin{figure}[th]
\centering
\includegraphics[width=0.65\linewidth]{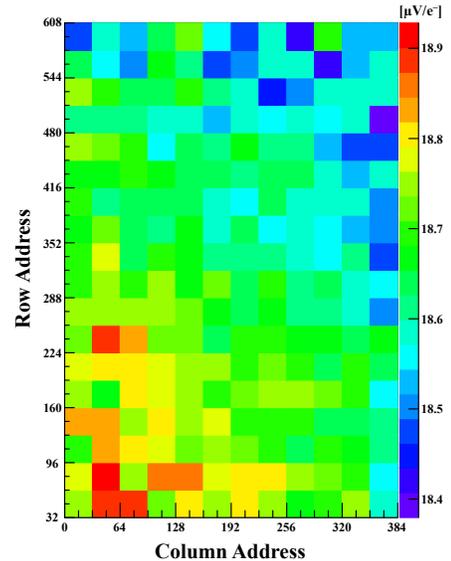}
\caption{Gain map of $32 \times 32$ pixel islands (cells) of XRPIX5b.} 
\label{fig:5b_gain_map}
\end{figure}

We evaluated the uniformity of the gain of XRPIX5b. 
Irradiating XRPIX5b with X-rays/$\gamma$-rays from a $^{57}{\rm Co}$ source, we obtained single-pixel events spectra from each $32 \times 32$ pixel island, which 
we refer to as cells hereafter, and calculated gains by analyzing each of the spectra. 
We applied a back-bias voltage of 200~V to the sensor layer of XRPIX5b for full depletion, and operated it in the frame readout mode. 
The sensor was cooled to $-60~^\circ{\rm C}$ to suppress the leakage current. 
The distribution of the gains is plotted in Fig.~\ref{fig:5b_gain_dist}, which indicates that the gain variation is at a $1\%$ level. 
We note that the statistical errors in the gains are $< 0.1~\mu{\rm V}/e^-$ and are negligible here. 
The measured cell-to-cell gain variation is not problematic for most applications. 
However, it is evident that the gains have a systematic tendency in the map shown in Fig.~\ref{fig:5b_gain_map}. 
We found that cells toward the upper-right corner tend to have lower gains. 
Since the signal output pad is located at the bottom-left corner of the device, we speculate that the tendency 
is due to the trace resistance of the signal lines.

\section{Conclusion}
We have been developing SOI pixel sensors for future X-ray astronomy missions such as FORCE. 
By introducing a new pixel structure called PDD, we have achieved an energy resolution of $\sim 150$~eV (FWHM) 
for 6.4-keV X-rays with our latest device, XRPIX6E. 
The energy resolutions obtained in the frame readout mode and the event-driven mode are comparable to each other, indicating 
that the capacitive coupling between the sensor and circuit layers is significantly reduced by the PDD structure. 
We processed a large-sized sensor named XRPIX5b, which has an imaging size of $21.9~{\rm mm} \times 13.8~{\rm mm}$. 
We demonstrated the imaging capability of XRPIX5b. 
We found that the gain uniformity of XRPIX5b is at a satisfactory level for practical use.

\end{document}